\documentclass[aps,showpacs,twocolumn,showkeys,amsmath,amssymb,floatfix,nofootinbib,preprintnumbers]
{revtex4}
\usepackage{bm}

\begin{document}
 
\title{Quarkonium $h$ States As Arbiters of Exoticity}

\author{Richard F. Lebed}
\email{richard.lebed@asu.edu}
\affiliation{Department of Physics, Arizona State University, Tempe,
Arizona 85287-1504, USA}

\author{Eric S. Swanson}
\email{swansone@pitt.edu}
\affiliation{Department of Physics and Astronomy, University of
Pittsburgh, Pennsylvania 15260, USA}

\date{May, 2017}

\begin{abstract}
  The mass splitting between the quarkonium spin-singlet state $h$
  ($J^{PC} = 1^{+-}$) and the spin average of the quarkonium
  spin-triplet states $\chi$ ($J^{PC} = 0^{++}, 1^{++}, 2^{++}$) is
  seen to be astonishingly small, not only in the charmonium and
  bottomonium cases where the relevant masses have been measured, but
  in positronium as well.  We find, both in nonrelativistic quark
  models and in NRQCD, that this hyperfine splitting is so small that it
  can be used as a test of the pure $Q\bar Q$ content of the states.
  We discuss the $2P$ states of charmonium in the vicinity of
  $3.9$~GeV, where the putative exotics $X(3872)$ and $X(3915)$ have
  been seen and a new $\chi_{c0}(2P)$ candidate has been observed at
  Belle.
\end{abstract}


\maketitle


\section{Introduction} \label{sec:Intro}

The spectrum of charmonium-like states in the region near 3.9~GeV is
exceptionally intricate and interesting.  In addition to containing
states that are believed to be the conventional $c\bar c$ $1{}^3 \!
D_1$ [$\psi (3770)$], $1{}^3 \! D_2$ [$\psi (3823)$], and $2{}^3 \!
P_2$ [$\chi_{c2}(2P)$]~\cite{Olive:2016xmw}, this range has produced
several unexpected states, including the most famous exotic candidate
$X(3872)$ ($J^{PC} = 1^{++})$, as well as the $0^{++}$ (or even
possibly $2^{++}$~\cite{Zhou:2015uva}) $X(3915)$, the $1^{+-}$ $Z_c^0
(3900)$ that is the neutral isospin partner of the $Z_c^+ (3900)$, and
the $X(3940)$, whose $J^{PC}$ remains unknown.  For a review of these
states and more, see Ref.~\cite{Lebed:2016hpi}.

Missing from this list are several expected states in the $2P$ band,
such as the conventional $0^{++}$ $\chi_{c0}(2P)$ and $1^{++}$
$\chi_{c1}(2P)$, and the $1^{+-}$ $h_c (2P)$.  Indeed, the $X(3872)$
has long been argued to have at least a substantial $\chi_{c1}(2P)$
component, while the $X(3915)$ was briefly listed by the Particle Data
Group as $\chi_{c0}(2P)$ until serious doubts were raised about this
identification (especially its lack of $D\bar D$ final
states)~\cite{Guo:2012tv,Wang:2014voa,Olsen:2014maa}; for example,
$X(3915)$ might even be the lightest $c\bar c s\bar s$
state~\cite{Lebed:2016yvr}.  The crucial importance of sorting out the
states in the $2P$ charmonium sector, in order to determine which
states are (mostly) exotic and which are not, was emphasized as a
central experimental goal in Ref.~\cite{Briceno:2015rlt}.  A very
recent attempt in this direction appears in Ref.~\cite{Yu:2017bsj}.

The latest chapter in this saga is the Belle
observation~\cite{Chilikin:2017evr} of a $\chi_{c0}(2P)$ candidate
decaying to $D\bar D$, with mass $3862^{+26+40}_{-32-13}$~MeV and
width $201^{+154+88}_{-67-82}$~MeV.  While these uncertainties are
quite large, the significance of the signal is substantial ($6.5
\sigma$).  With the $\chi_{c2}(2P)$ and hopefully the $\chi_{c0}(2P)$
now in hand, one can at last begin a serious study of mass splittings
within this multiplet, with an eye toward testing the expectations for
pure $c\bar c$ composition versus mixing with multiquark or hybrid
components.

The mass difference of interest in this paper is the hyperfine
splitting between the quarkonium spin-singlet state $h$ (${}^1 \! P_1,
\, J^{PC} \! = 1^{+-}$) and the spin average of the quarkonium
spin-triplet states $\chi$ (${}^3 \! P_{0,1,2}, \, J^{PC} \! = 0^{++},
1^{++}, 2^{++}$):
\begin{equation} \label{eq:DeltaDef} \Delta \equiv M_h - \frac{1}{9}
  \left[ 1 \! \cdot \! M_{\chi_0} + 3 \! \cdot \! M_{\chi_1} + 5 \!
    \cdot \! M_{\chi_2} \right] \, .
\end{equation}
A complete set of experimental data for determining $\Delta$ is
currently available in only 4 cases: $2P$ positronium (which, in the
standard notation for positronium, is the lowest $P$
wave)~\cite{Hagena:1993,Ley:1994}, $1P$ charmonium, and $1P$ and $2P$
bottomonium~\cite{Olive:2016xmw}.  The corresponding $\Delta$ values
are presented in Table~\ref{tab:DeltaVals}.  In every case, the value
of $\Delta$ is zero to within experimental uncertainties, making the
tight relationship among $P$-level states highly predictive for cases
(such as $2P$ charmonium) in which some of the states have not yet
been observed.
\begin{table*}
  \caption{Experimental values of $\Delta$ in MeV for quarkonium and in
    MHz for positronium.  For quarkonium the state masses entering
    Eq.~(\ref{eq:DeltaDef}) are listed, while for positronium the
    differences $(2{}^3 \! S_1 - 2{}^{2S+1} \! P_J)$ are presented.}
\label{tab:DeltaVals}
\begin{tabular}{c|ccccc}
  \hline\hline
  System & $h ({}^1 \! P_1)$ & $\chi_{0} ({}^3 \! P_0)$ & $\chi_{1}
  ({}^3 \! P_1)$ & $\chi_{2} ({}^3 \! P_2)$ & $\Delta$ \\
  \hline
  $c\bar c(1P)$ & 3525.38(11) & 3414.75(31) & 3510.66(7) & 3556.20(9)
  & +0.08(13) \\
  $c\bar c(2P)$ & $-$ & $3862^{+26+40}_{-32-13}$ & $-$ & 3927.2(2.6)
  & $-$ \\
  $b\bar b(1P)$ & 9899.3(8) & 9859.44(42)(31) & 9892.78(26)(31) &
  9912.21(26)(31) & $-0.57(88)$ \\
  $b\bar b(2P)$ & 10259.8(1.2) & 10232.5(4)(5) & 10255.46(22)(50) &
  10268.65(22)(50) & $-0.44(1.31)$ \\
  Ps & 11180(5)(4) & 18499.65(1.20)(4.00) & 13012.42(67)(1.54) &
  8624.38(54)(1.40) & +4.31(6.50) \\
\hline\hline
\end{tabular}
\end{table*}

In this short paper we explore the physical reason for this remarkable
relationship in quarkonium; indeed, $\Delta$ is so small that one may
call it an {\it ultrafine\/} splitting.  We then show how it may be
applied to the confusing set of charmonium states around $3.9$~GeV to
uncover an unambiguous signal of exoticity, by which we mean a
non-$c\bar c$ state component.  Finally, we remind the reader of the
ongoing and proposed experiments designed to uncover missing
quarkonium states.

This paper is organized as follows.  In Sec.~\ref{sec:Operators} we
identify and discuss the operators potentially contributing to the
``ultrafine'' mass difference Eq.~(\ref{eq:DeltaDef}) and related
combinations.  Section~\ref{sec:QuarkPot} identifies the origin of the
relevant operator in quark potential models and explains the origin of
its numerical suppression; Sec.~\ref{sec:NRQCD} does the same for the
non-relativistic QCD effective theory.  In Sec.~\ref{sec:nonpert} we
discuss the nonperturbative heavy-quark limit and the effect of the
appearance of partonic degrees of freedom beyond the heavy
quark-antiquark pair, and in Sec.~\ref{sec:Exotics} describe the use
of the ``ultrafine'' relation in identifying the presence of exotic
(non-$Q\bar Q$) components in the candidate states and the prospects
for observing the missing states.  Section~\ref{sec:Concl} summarizes
and concludes.

\section{Operators Contributing to $\Delta$} \label{sec:Operators}

The hyperfine interaction is defined, as usual, as a direct coupling
between the intrinsic spins of the component fermions of the state.
In the case of $f\bar f$ bound states, where $f$ is a spin-$\frac 1 2$
fermion, one can produce only a finite number of linearly independent
operators contributing to the mass from the basic ingredients of
quark-spin ${\bf S}_f$, ${\bf S}_{\bar f}$ and orbital angular
momentum ${\bf L}$ operators.  For example, a quark-spin operator that
transforms under an irreducible representation with spin greater than
two cannot, by the Wigner-Eckart theorem, contribute to matrix
elements of states containing only two spin-$\frac 1 2$ quarks.  On
the other hand, operators sensitive to arbitrarily high powers of
squared quark momenta (but no spin dependence) might be generated by
the fine details of quark distributions within the hadron, but their
contributions to hadron masses are proportional to those arising from
any spin-symmetric operator, such as the quark-mass operator.

To put the discussion on a firm footing, we define the usual operators
in configuration space:
\begin{eqnarray}
\label{eq:hyperfine}
{\bf S}_f \! \cdot {\bf S}_{\bar f} && {\rm (hyperfine)} \, , \\
\label{eq:spinorbit}
{\bf S} \cdot {\bf L} && {\rm (\mbox{spin-orbit})} \, , \\
\label{eq:tensor}
\overset{\text{\tiny$\bm\leftrightarrow$}}{T}
\equiv ({\bf S}_f \! \cdot \hat {\bf r})
({\bf S}_{\bf f} \! \cdot \hat {\bf r})
-\frac 1 3 {\bf S}_f \! \cdot {\bf S}_{\bar f} && {\rm (tensor)} \, ,
\end{eqnarray}
where ${\bf S} \equiv {\bf S}_f \! + {\bf S}_{\bar f}$.  The operators
can be expressed just as easily in momentum space by replacing
$f$-$\bar f$ relative position operator ${\bf r}$ with the relative
momentum operator ${\bf q}$ and replacing {\bf L} with ${\bf q} \times
{\bf p}$, where ${\bf p}$ is the total momentum operator.  In any
case, these are the only three independent spin-dependent operators
that arise up to quadratic order in ${\bf S}_{f,\bar f}$; and since
all linearly independent operators arising beyond quadratic order
transform as spin greater than two, the list in
Eqs.~(\ref{eq:hyperfine})--(\ref{eq:tensor}) is complete.  For
example, the operator $({\bf S} \! \cdot {\bf L})^2$ can be
shown\footnote{This fact provides one convenient
method~\cite{Landau:1977} for calculating matrix elements of
$\overset{\text{\tiny$\bm\leftrightarrow$}}{T}$, such as those given
in Eq.~(\ref{eq:Tensor}).}  for any given multiplet of $f\bar f$
states to be linearly dependent on the ones above plus the operator
${\bf S}^2 {\bf L}^2$.

Now consider any multiplet of $f\bar f$ states that, in the language
of a quark potential model, carry the same principal quantum number
$n$ and orbital angular momentum $L$.  For states with $S = 0$ [and
hence $J = L$: quarkonium $\eta$ and $h$], matrix elements of both the
spin-orbit and tensor operator vanish by the Wigner-Eckart theorem
since the operators transform as $S=1$ and $S=2$, respectively.  The
same matrix elements vanish for all $L=0$ states, so one immediately
sees that any $n{}^3\!  S_1$-$n{}^1 \! S_0$ hyperfine splitting---in
which all spin-independent mass terms cancel---depends only on the
hyperfine operator, as one might expect.  For $L>0$ and $S=1$,
$J=L-1$, $L$, and $L+1$ are allowed, and using the results
\begin{equation}
\langle {\bf S} \! \cdot {\bf L} \rangle = \frac 1 2 \left[ J(J+1) -
  L(L+1) - S(S+1) \right] \, ,
\end{equation}
and
\begin{equation}
\left< \overset{\text{\tiny$\bm\leftrightarrow$}}{T} \right> = \left\{
\begin{array}{rl}
-\frac{L+1}{6(2L-1)} \, , & J = L-1 \, , \\
+\frac 1 6           \, , & J = L   \, , \\
-\frac{L}{6(2L+3)}   \, , & J = L+1 \, ,
\end{array} \right. \label{eq:Tensor}
\end{equation}
one can quickly check that adding these matrix elements weighted by
the $2J+1$ degenerate spin states for each level gives a vanishing
result.  In other words, the spin-averaged matrix elements of any trio
of spin-triplet states $n{}^3 \! L_{J = L-1}$, $n{}^3 \! L_{J = L}$,
$n{}^3 \! L_{J = L-1}$ with orbital angular momentum larger than zero
vanish for the spin-orbit and tensor operators.  The reason is not so
mysterious: Although expressed in the $\left| J, J_z, L, S=1 \right>$
basis, the states form a complete multiplet in the $\left| L, L_z,
S=1, S_z \right>$ basis, while the spin-orbit and tensor operators,
being irreducible operators of rank greater than zero, are traceless.
On the other hand, all of these states have the same spin-independent
mass terms, which is also the same as that of the corresponding
spin-singlet $n{}^1 \! S_{J=L}$.  In total, the mass combination for
orbital angular momentum larger than zero defined by:
\begin{eqnarray}
\Delta_{n,L} & \equiv & M(n{}^1 \! L_{J=L}) \nonumber \\
& - & \frac{2L-1}{3(2L+1)} M(n{}^3 \! L_{J = L-1}) \nonumber \\
& - & \frac{2L+1}{3(2L+1)} M(n{}^3 \! L_{J = L}) \nonumber \\
& - & \frac{2L+3}{3(2L+1)} M(n{}^3 \! L_{J = L+1}) \, ,
\label{eq:HyperGen}
\end{eqnarray}
of which Eq.~(\ref{eq:DeltaDef}) is simply the case $L=1$ for
quarkonium, receives contributions only through the hyperfine
operator.

All mass combinations $\Delta_{n,0} \equiv M(n{}^1 \! S_0) - M(n{}^3
\! S_1)$ and $\Delta_{n,L}$ of Eq.~(\ref{eq:HyperGen}) for $L > 0$ are
thus pure hyperfine splittings.  In order to see why the latter
deserve the label ``ultrafine,'' we consider the origin of the
hyperfine operator in three useful dynamical formalisms:
quark-potential models, nonrelativistic QCD (\mbox{pNRQCD}), and the
heavy-quark expansion of QCD, including nonperturbative effects.

\section{Quark Potential Models} \label{sec:QuarkPot}

The form of the operator multiplying ${\bf S}_f \! \cdot {\bf S}_{\bar
f}$ is clearly crucial for determining the relative size of various
hyperfine interactions.  In the case of electromagnetic interactions,
the direct spin-spin coupling is pointlike, being proportional to the
wave function of the state at zero spatial separation {\bf r} between
the spins, a fact first noted by Fermi~\cite{Fermi:1930}.  Such a term
is proportional to $\delta^{(3)}({\bf r})$, and arises naturally in
the nonrelativistic reduction of terms in the Dirac equation (and more
generally, QED) describing the interaction of two charged particles,
known as the Breit Hamiltonian~\cite{Breit:1932}.  In that context, it
appears through the Laplacian operator acting upon the Coulomb
potential $1/r$.  Of course, this $1/r$ simply occurs as the Fourier
transform of the momentum-space propagator $1/q^2$ of the massless
photon.

The equivalent Breit Hamiltonian for the case of potential
interactions in quark systems, as applied to hadron masses, was
expressed in De~Rujula, Georgi, and Glashow~\cite{DeRujula:1975qlm}.
Since QCD also has massless gauge bosons in the form of gluons, one
finds that the corresponding spin-spin term is proportional to
$\delta^{(3)}({\bf r})$ in the short-distance limit in which the
interaction is dominated by one-gluon exchange.  {\em Any\/} quark
potential model in which the potential $V(r)$ contains a piece
representing one-gluon exchange will exhibit this feature.  This term
is represented for example, in the most thorough recent analysis of
charmonium masses~\cite{Barnes:2005pb} as a Gaussian of width
$1/\sigma$:
\begin{equation} \label{eq:smear}
\delta^{(3)}({\bf r}) \to \tilde \delta_\sigma (r) \equiv
\left( \frac{\sigma}{\sqrt{\pi}} \right)^3 e^{-\sigma^2 r^2} \, .
\end{equation}
``Smearing'' of this sort is necessary to regularize the delta
function (thereby making it a well-defined three-dimensional
quantum-mechanical operator), and because nonzero hyperfine splitting
is evident in the spectrum for radially excited $S$-wave states.

The most interesting feature of the $\delta^{(3)}({\bf r})$ dependence
is that it is only supported by wave functions that are nonvanishing
at the origin. Of course, one well-known feature of quantum mechanics
is that wave functions with orbital angular momentum quantum number
$L$ scale as $r^{L}$ near the origin.  Therefore, one naturally
expects the $S$-wave hyperfine splittings to be numerically much
larger than those with $L>0$, hence the term ``ultrafine.''  The
evidence from Table~\ref{tab:DeltaVals} strongly supports this
conclusion; for example, the $S$-wave hyperfine splitting in the $n=1$
level of charmonium is $m_{J/\psi} - m_{\eta_c} = 113.5(5)$~MeV\@.
One anticipates that the $D$-wave splittings would be even smaller
than those in the $P$-wave.

Using the smearing function of Eq.~(\ref{eq:smear}) and the parameter
value~\cite{Barnes:2005pb} $\sigma = 1.0946$~GeV, we find values of
$\Delta_{n,1}$ of order 3--10~MeV, which are much larger than those
observed.  This result is a reflection of the relatively small value
of $\sigma$ used in the model, which is driven by the much larger
observed $S$-wave hyperfine splittings, $\Delta_{n,0}$.


\section{NRQCD} \label{sec:NRQCD}

In the heavy-quark limit of QCD, all operators dependent upon the
heavy-quark flavor or spin are suppressed by powers of the heavy-quark
mass $m_Q$, and the relevant finite dynamical parameter becomes the
heavy-quark four-velocity $v$.  Effects at energy scales higher than
$m_Q$ are integrated out in the usual Wilsonian fashion, leading to
the {\it heavy-quark effective theory}~\cite{Manohar:2000dt}.
Spin-spin and tensor operators, containing two heavy-quark spin
operators, are therefore suppressed by $1/m_Q^2$.

When more than one heavy quark is present, as in quarkonium, new
scales arise, and their modes must be integrated out successively: in
decreasing magnitude, these are the ``hard'' scale $m_Q$ (leading to
the effective theory of {\it non-relativistic QCD\/}
(NRQCD)~\cite{Caswell:1985ui,Bodwin:1994jh}, the ``soft'' scale $m_Q
v$, and the ``potential'' scale (energies $\sim m_Q v^2$, momenta
$\sim m_Q v$), leading to the effective theory called
\textit{potential non-relativistic QCD}
(pNRQCD)~\cite{Pineda:1997bj,Brambilla:1999xf}.  The remaining modes
are ``ultrasoft'' (energies and momenta $\sim m_Q v^2$).

The state-of-the-art calculations in this program are now performed at
next-to-next-to-next-to leading order (NNNLO) in NRQCD; the
corresponding Hamiltonian was obtained in Ref.~\cite{Kniehl:2002br}
and the heavy-quarkonium spectrum for states of arbitrary quantum
numbers, including terms of $O(m_Q \alpha_s^5 \ln \alpha_s)$, was
presented in Ref.~\cite{Brambilla:1999xj}.  The corresponding
expressions including $O(m_Q \alpha_s^5)$ terms, which we use here,
appear in Ref.~\cite{Kiyo:2014uca}.

Using the NNNLO results from Ref.~\cite{Kiyo:2014uca}, one computes
the $1S$ hyperfine splitting:
\begin{eqnarray}
\lefteqn{\Delta_{1,0} = \frac{1}{3} m_Q \alpha_s^4 C_F^4} &&
   \nonumber \\
   &&-  \frac{m_Q \alpha_s^5 C_F^4}{108\pi} \bigg[
   6 \pi ^2 \beta_0 - 72 \beta_0 + 20 C_A + 18 C_F \nonumber \\
   && + 63 C_A \log \left(\frac{1}{\alpha_s C_F} \right)
   -72  \beta_0 \log \left( \frac{\mu }{\alpha_s C_F m_Q} \right)
   \nonumber \\
   && + 8 T_F  n_\ell -54 T_F + 54 T_F \log 2 \bigg] \, ,
   \label{eq:1SHyper}
\end{eqnarray}
where $\beta_0 = (11C_A - 2n_\ell)/3$, with $n_\ell$ being the number
of light fermion species appearing in loop corrections ({\it i.e.}, as
short-distance degrees of freedom only) and $\mu$ being the
renormalization point.  The usual color and trace factors $C_A = N_c
\to 3$, $C_F = (N_c^2 - 1)/2N_c \to \frac 4 3$, and $T_F = \frac 1 2$
also appear.  It is worth noting that the leading $[O(\alpha_s^4)]$
term in this expression gives only 13~MeV for $\alpha_s = 0.3$ and
$m_c = 1.5$~GeV -- too low by a factor of nearly 9 compared to the
experimental value given above; however, the $O(\alpha_s^5)$ term not
only exhibits a strong dependence upon $\mu$ but turns out to be of
the same numerical order as the leading term.  One concludes that the
splitting $\Delta_{1,0}$ is not yet under control in the NRQCD
result Eq.~(\ref{eq:1SHyper}).

In comparison, the ultrafine combination at this order in NRQCD
computed using Ref.~\cite{Kiyo:2014uca} is much simpler:
\begin{equation} \label{eq:DeltaQuark}
\Delta_{n,1} = \frac{m_Q C_F^4 \alpha_s^5}{432 \pi (n+1)^3}
\left( 8 T_F n_\ell -
C_A \right) \, .
\end{equation}
This expression is smaller both parametrically [by a power of
$\alpha_s (m_Q)$] and numerically (by the large denominator factor)
than the usual hyperfine splitting.

The origin of the extra suppression, leading to the ``ultrafine''
label, arises for precisely the same algebraic reason as it does for
the nonrelativistic quark potential model: The same set of
spin-dependent operators as in
Eqs.~(\ref{eq:hyperfine})--(\ref{eq:tensor}) arises in NRQCD, and the
spin-spin operator Eq.~(\ref{eq:hyperfine}) again appears with the
contact interaction coefficient $\delta^{(3)}({\bf
  r})$~\cite{Kniehl:2002br}.  The $L > 0$ overlap integrals are again
severely suppressed, and indeed only survive due to the
renormalization of the $\alpha_s$ coefficient to
Eq.~(\ref{eq:hyperfine}) to include terms of the form $\ln(r)$ that
are singular for $r \to 0$.  One thus obtains the ultrafine
suppression in effective field theory treatments of QCD.

The values obtained from Eq.~(\ref{eq:DeltaQuark}) are remarkably
small: One obtains 9.5~keV, 2.8~keV, 3.8~keV, and 1.1~keV for 1P and
2P charmonium, and 1P and 2P bottomonium, respectively; such
differences are far smaller than the central values given in
Table~\ref{tab:DeltaVals}.  The positronium result is expected to be
even further suppressed, to $O(\alpha^6)$~\cite{Lamm:2017}.

In comparison, the Schnitzer ratio~\cite{Schnitzer:1976td},
\begin{equation}
R_1 \equiv \frac{M({}^3 \! P_2) - M({}^3 \! P_1)}
{M({}^3 \! P_1) - M({}^3 \! P_0 )} \, ,
\end{equation}
is known to be exactly $\frac 4 5$ at leading order in $\alpha_s$, and
this result is borne out in NRQCD calculations~\cite{Kiyo:2014uca}.
However, the $O(\alpha_s)$ corrections to this ratio have a strong
$\mu$ dependence, which can be used to accommodate its rather
different experimental value for $1P$ charmonium, $\sim 0.47$.  One
may anticipate a similar strong $\mu$ dependence to arise in the next
(uncomputed) $O(\alpha_s^6)$ corrections to
Eq.~(\ref{eq:DeltaQuark}), but one still expects the measured values
for $\Delta_{n,1}$ to remain no larger than $O(10 \, {\rm keV})$.

\section{Nonperturbative Heavy-Quark Limit}
\label{sec:nonpert}

The matrix elements that appear in the NRQCD results reported in the
previous section are evaluated in the heavy-quark limit, for which
nonrelativistic Coulombic meson wave functions are appropriate.  This
approximation breaks down as the quarks become lighter and the scale
$\Lambda_{\rm QCD}$ becomes more relevant.  Furthermore, the
nonperturbative regime also becomes more relevant as the radial
quantum number increases, because larger spatial scales are probed.
Under these conditions, mass splittings that had been proportional to
the heavy quark mass can scale as $\Lambda^3_{\rm QCD}/m_Q^2$.  It is
thus prudent to enquire into the regime of validity of the NRQCD
computations presented above.

An analogous problem arises in the application of the operator-product
formalism to the interaction of heavy mesons with hadronic matter.  In
this case,
Peskin~\cite{Peskin:1979va,Luke:1992tm,Gottfried:1977gp,Goldberg:1975bz}
has estimated that the method is reliable if:
\begin{equation}
  m_Q \gg n^2 \frac{\Lambda_{\rm QCD}}{\alpha_s^2(r_Q^{-1})} \, .
\end{equation}
It was subsequently argued that this expression should contain a
numerical coefficient of order 10~\cite{Lakhina:2003pj}.  As a result,
the operator-product expansion (in this context) is never valid for
physical quarks.

In view of this issue, we seek to estimate the nonperturbative
behavior of the hyperfine matrix element that contributes to the
hyperfine and ultrafine splittings.  A formalism for examining this
question was developed long ago by Eichten and
Feinberg~\cite{Eichten:1980mw}, who applied the heavy-quark expansion
to the Wilson loop to obtain expressions for the spin-dependent
interaction of heavy quarks at order $1/m_Q^2$.  The result for the
coefficient of the spin-spin term is proportional to the temporal
integral of the matrix element of chromomagnetic fields.
Subsequently, a somewhat more transparent, but equivalent, expression
was obtained with a Foldy-Wouthuysen reduction of the QCD Hamiltonian
in Coulomb gauge~\cite{Szczepaniak:1996tk}:
\begin{eqnarray}
  \lefteqn{V_{\rm hyp}^{(n^{\vphantom\ell}_R)}
  ({\bf R}={\bf r}_Q-{\bf r}_{\bar Q})
    = \alpha_s \frac{4\pi}{3 m_Q^2}\, {\bf S}_f 
    \! \cdot {\bf S}_{\bar f}} && \nonumber \\ 
  & \times & \sum_{m \neq n} \frac{1}{\epsilon_n(R) -
    \epsilon_m(R)} \nonumber \\
  & \times & \left< n_R; {\bf r}_Q , {\bf r}_{\bar Q} \left|
    \int d^3x \, h^\dagger({\bf x}) {\bf B}({\bf x}) h({\bf x})
    \right| m_R; {\bf r}_Q , {\bf r}_{\bar Q} \right> \nonumber \\
  & \cdot & \left< m_R; {\bf r}_Q , {\bf r}_{\bar Q} \left|
    \int d^3y \, \chi^\dagger({\bf y}) {\bf B}({\bf y}) \chi({\bf y})
    \right| n_R; {\bf r}_Q , {\bf r}_{\bar Q} \right>
    \nonumber \\
  & + & (h \leftrightarrow \chi) \, .
\label{eq:V4}
\end{eqnarray}
States are labeled with the coordinates of the static quarks, ${\bf
r}_Q$, ${\bf r}_{\bar Q}$, and gluonic quantum numbers are denoted by
$m_R, n_R$.  Heavy-quark and -antiquark creation operators are labeled
by $h^\dagger$ and $\chi^\dagger$, respectively.  The operators {\bf
B} are chromomagnetic fields.  One of the fields is evaluated at the
position of the quark ${\bf r}_Q$, while the other is evaluated at the
antiquark position ${\bf r}_{\bar Q}$.  Evaluating both fields on a single quark
line is possible, but does not yield a spin-dependent interaction.

Perturbatively, the matrix element of Eq.~(\ref{eq:V4}) becomes
\[
V_{\rm hyp}^{(n_R)}(R) \propto \nabla_Q \cdot \nabla_{\bar Q} \,
\langle {\bf A} ({\bf r}_Q,t) \! \cdot {\bf A} ({\bf r}_{\bar Q},t')
\rangle \, .
\]
The matrix element is proportional to $1/R$, and the hyperfine
interaction is then proportional to $\delta^{(3)}({\bf R})$.  A simple
nonperturbative estimate can be made by introducing a gluon effective
mass term into the gluon propagator.  In this case, one obtains a
Yukawa-type interaction, with a range given by the inverse gluon
effective mass, which is also seen to be short-ranged.

More generally, one can argue that infinitely many strongly
interacting virtual gluons tend to decorrelate the chromomagnetic
fields rapidly as the interquark separation is
increased~\cite{Szczepaniak:1996tk}.  This expectation is confirmed in
quenched lattice computations of the chromomagnetic field correlation
in the presence of a Wilson loop~\cite{Koma:2006fw}, in which it is
found that a Yukawa potential with a gluon mass of approximately
2.5~GeV fits the simulation well.  The potential itself is zero,
within statistical error, for $R > 0.2$~fm.


\section{Ultrafine Splittings and Exotics} \label{sec:Exotics}

All of the previous discussion leads to the same conclusion: The
heavy-quark hyperfine interaction is short-ranged.  Thus, matrix
elements of the interaction must decrease with radial and orbital
quantum number.  Furthermore, experiment indicates that
$\Delta_{1,1}(c\bar c)$, $\Delta_{1,1}(b\bar b)$, and $\Delta_{2,1}(b
\bar b)$ are all small, and hence this quantity must be small for all
$n$ and $L$ in the charmonium and bottomonium systems.

This conclusion follows from the quark model, which does not consider
coupled channels; NRQCD, which only considers the short-range
contribution of light quarks; heavy-quark QCD, which suppresses the
effect of light quarks; or quenched lattice computations.  Thus it is
possible that long-distance light-quark effects -- such as those
manifested in meson-meson contributions to quarkonium states, or by
$Q\bar Q q \bar q$ wave function components -- can ruin the
relationship $\Delta_{n,L} \ll \Lambda_{\rm QCD}$.  But this condition
can be taken as the {\em definition\/} of a crypto-exotic state that
contains important light-quark degrees of freedom.

In view of this observation and the putative exotic nature of the
$X(3872)$, we suggest that measuring the mass of the $h_{c}(2P)$ and
computing $\Delta_{2,1}(c\bar c)$ will unambiguously reveal if the
$X(3872)$ contains important light-quark degrees of freedom.  This
conclusion, of course, relies upon assuming that $\chi_{c2}(2P)$, the
new $\chi_{c0}(2P)$ candidate, and the undiscovered $h_c (2P)$ are
pure $c\bar c$ states; at minimum, one can conclude that a substantial
violation of the relation $\Delta_{n,L} \ll \Lambda_{\rm QCD}$ in
heavy quarkonium points to {\em at least one\/} of the states
containing a significant non-$Q\bar Q$ component.  Indeed, even an
effect giving $\Delta_{n,L} > O(\Lambda_{\rm QCD}^3 / m_Q^2)$, the
heavy-quark spin-symmetry expectation, will warrant close attention.

The same comments hold for the $D$-wave $c \bar c$ states: the $\psi
(3770)$ and $\psi (3823)$ are believed to be $1{}^3 \! D_1$ and $1{}^3
\! D_2$, respectively; the observation of a spin-0 $\eta_c$ $1{}^1 \!
D_1$ or spin-3 $\psi$ $1{}^3 \! D_3$ will allow a precise prediction
of the mass of the other, while a measurement of both will allow one
to test for a non-$c\bar c$ component in this multiplet.

The prospects for experimentally measuring the $2P$ charmonium
ultrafine splitting are encouraging.  For example, \mbox{BESIII}
observed the $Z_c^0(4020)$ in the reaction $e^+e^- \to \pi \pi
h_c(1P)$~\cite{Ablikim:2013wzq}.  A similar effort could yield a
signal for $\pi\pi h_c(2P)$, with the $h_c(2P)$ being detected in its
$D \bar D^*$ decay mode~\cite{rem}.

Alternatively, attempting to find $\chi_{cJ}(2P)$ with $J>0$ in the
recoil mass products $X$ of $e^+e^- \to J/\psi X$ is not expected to
be profitable, since this channel has been seen to be dominated by
$\eta_c(1S)$, $\eta_c(2S)$, $\chi_{c0}(1P)$, and the $X(3940)$, with
little evidence for $\chi_{c1}(1P)$, $\chi_{c2}(1P)$, or any of the
expected $\chi_{cJ}(2P)$~\cite{Abe:2007jna}.

Examining the process $B \to K \omega J/\psi$ should shed light on the
enigmatic $X(3915)$, which likely plays a role in the charmonium $2P$
spectrum. Finally, collecting sufficient data in $B \to K D \bar D^*$
should permit observation of the $h_c(2P)$ and the
$\chi_{c1}(2P)$~\cite{so}.

\section{Conclusions} 
\label{sec:Concl}

We have argued that the splitting $\Delta_{n,L}$ defined in
Eq.~(\ref{eq:HyperGen}) is robustly ``ultrafine'' in the absence of
explicit long-distance light-quark degrees of freedom, and therefore
can serve as an unambiguous test of the ``coupled-channel exoticity''
of quarkonium states.  Prospects for applying this test in the
charmonium $2P$ sector appear to be good.

It is interesting to speculate on the applicability of this idea to
quarkonium hybrid states.  The chromomagnetic matrix element of
Sec.~\ref{sec:nonpert}, which gives the direct quark-antiquark spin
coupling ${\bf S}_f \! \cdot {\bf S}_{\bar f}$, depends upon the
gluonic state of the heavy-quark meson [which is explicit in
Eq.~(\ref{eq:V4}) and is implicit in the Eichten-Feinberg formalism].
Naively, the ultrafine splitting could be large in states with
substantial hybrid components. However, the arguments of
Sec.~\ref{sec:nonpert} lead us to believe that this will not be the
case, because the addition of valence gluonic degrees of freedom
should not reverse the rapid decorrelation of the chromomagnetic
fields. This expectation can be checked directly with lattice
measurements of $V_{\rm hyp}^{(\rm hybrid)}$ [Eq.~(\ref{eq:V4})],
which should be readily achievable with present capabilities.  The
application of this result to splittings in complete hybrid multiplets
will be complicated by the addition of many new spin-dependent
operators in the heavy-quark expansion, but should, nevertheless, also
be of interest.

\begin{acknowledgments}
  \vspace{-2ex} We gratefully acknowledge H.~Lamm, R.E.~Mitchell, and
  S.~Olsen for their expert advice.  This work was supported by the
  National Science Foundation under Grant No.\ PHY-1403891.
\end{acknowledgments}


\end{document}